\documentclass[preprint,showpacs,prd]{revtex4}%
\usepackage{amsmath}
\usepackage{amsfonts}
\usepackage{amssymb}
\usepackage{graphicx}

\def\be{\begin{equation}}
\def\ee{\end{equation}}
\def\bea{\begin{eqnarray}}
\def\eea{\end{eqnarray}}

\begin{document}

\title{Geometrothermodynamics in Ho\v rava--Lifshitz gravity }

\author{Hernando  Quevedo$^{1,2}$, Alberto S\'anchez$^3$, Safia Taj$^{2,4}$, and Alejandro V\'azquez$^5$}
\email{quevedo@nucleares.unam.mx, asanchez@nucleares.unam.mx, safiataaj@gmail.com, alec_vf@nucleares.unam.mx}
\affiliation{
$^1$Instituto de Ciencias
Nucleares, Universidad Nacional Aut\'onoma de M\'exico,
 AP 70543, M\'exico, DF 04510, Mexico\\
$^2$ICRANet, Dipartimento di Fisica, Universit\`a di Roma "La
Sapienza",  I-00185 Roma, Italy\\
$^3$Departamento de
Posgrado, CIIDET, AP 752, Quer\'etaro, QRO 76000, Mexico\\
$^4$Center for Advanced
Mathematics and Physics, National University of Sciences and
Technology, H-12, Islamabad, Pakistan\\
 $^5$Facultad de Ciencias, 
Universidad Aut\'onoma del Estado de Morelos, 
Cuernavaca, MO 62210, Mexico}

\date{\today}

\begin{abstract}
We investigate the thermodynamic geometries of the most general 
static, spherically symmetric, topological black holes of 
the Ho\v rava--Lifshitz gravity. In particular, we show that
 a Legendre invariant metric derived in the context of geometrothermodynamics
 for the equilibrium manifold 
reproduces correctly the phase transition structure of these black holes.
Moreover, the limiting cases in which the mass, the entropy or the Hawking
temperature vanish are also accompanied by curvature singularities which
indicate the limit of applicability of the thermodynamics and the geometrothermodynamics
of black holes. The Einstein limit and the case of a black hole with flat horizon 
are also investigated. 

{\bf Keywords:} Ho\v rava-Lifshitz gravity, Black holes, Geometrothermodynamics, 
Phase transitions 
\end{abstract}

\pacs{04.70.Dy, 02.40.Ky}

\maketitle

\section{Introduction}
\label{sec:int}

General relativity and quantum mechanics are considered as
the cornerstones of modern physics; however, all the attempts to formulate a
unified theory of quantum gravity have been so far unsuccessful.
Interesting technical results have been obtained in different
approaches, but the physical problem remains open due, in part, to
the fact that general relativity turned out to be non-renormalizable
in the ultraviolet (UV) regime.

A field theoretical model which can be interpreted as a complete
theory of gravity in the UV limit was recently proposed by
Ho\v rava \cite{Horawa}. The model is renormalizable
\cite{Orla} and non-relativistic in the UV regime. Moreover, it reduces to
Einstein's gravity theory with a cosmological
constant in the infrared (IR) limit. Since space and time have
different scalings at the UV fixed point, i. e., $x^i \to  l x^i,\ t
\to l^z t$ where $z$ is the scaling exponent, the model is
usually named Ho\v rava-Lifshitz (HL) theory in the
literature \cite{edue}; 
in particular this theory is renormalizable if $z = 3$.

It was found that the Schwarzschild--anti de Sitter black hole solution is not recovered in the IR
limit, although Einstein's theory with cosmological model was
obtained at the level of the action \cite{BlackHoles}. This
difficulty was solved by introducing an additional parameter which
modifies the IR behavior \cite{GeneralizzazioneHorawa,adddreee}. 

The study of black holes and their thermodynamic properties in the HL theory has 
been the subject of intensive research in the last years. In particular, the geometry 
of the thermodynamics of black holes has been considered in several works 
\cite{bischak10,wei10,jjk10,ccs10} by using different approaches. A first approach 
consists in introducing in the space of equilibrium states 
the Weinhold \cite{Weinhold} metric which is defined as the Hessian of the internal energy $U$ 
of the system
\be
g_{ij}^W =\partial_i\partial_j U(S,N^r)\ ,
\label{wei}
\ee 
where $S$ is the entropy and $N^r$ represents the remaining extensive thermodynamic variables of 
the system. An alternative metric was proposed by Ruppeiner \cite{Ruppeiner,Ruppeiner2} as minus the 
Hessian of the entropy 
\be
g^R_{ij}= - \partial_i \partial_j S(U,N^r) \ ,
\label{rup}
\ee
and is related to the Weinhold metric through the line element relationship $ds^2_W = T ds^2_R$, where $T$ 
denotes the temperature. Both metrics have been applied to study the geometry of the thermodynamics of
ordinary systems \cite{ord,ord2,ord3,ord4,ord5,ord6,ord7,jjkb,jjka}; however, several inconsistencies and contradictions have been found 
in particular in the study of black hole thermodynamics \cite{am,ama,aman,scws,cc,sst,med,mz,hernando2}.
More recently, Liu, L\"u, Luo and Shao \cite{chinosmet} proposed to use the metric 
\be
g_{ij}^{LLLS} = \partial_i\partial_j \tilde U (N^k)\ ,
\label{llls}
\ee
where $\tilde U$ is any of the thermodynamic potentials that can be obtained from $U$ by means of a Legendre transformation.
In the special cases $\tilde U = U$ and $\tilde U = S$, one obtains the Weinhold and Ruppeiner metrics, respectively.

The inconsistencies that appear from using the above metrics are explained in the framework 
of geometrothermodynamics (GTD) \cite{quev07} as due to the fact that all those metrics are not invariant with respect
to Legendre transformations. In this work, we will use a Legendre invariant metric in the context of GTD to formulate
an invariant geometric representation of the thermodynamics of one of the most general black hole solutions 
known in the HL theory. We also compare our results with those obtained by using the Weinhold and Ruppeiner geometries.

This paper is organized as follows. In Sec. \ref{sec:hor} we review 
the main thermodynamic properties of the topological black hole solutions  
of the HL gravity.
In Sec. \ref{sec:gtd}, we review the formalism of GTD and discuss the non-invariance
of some of the known thermodynamic metrics.  Section \ref{sec:geo} contains 
the results of analyzing the thermodynamics of the topological black hole
solution using the Weinhold and Ruppeiner geometries. In Sec. \ref{sec:gtd1}, 
we investigate the geometrothermodynamics of the topological black hole and show that 
it agrees with the results following from the analysis of the corresponding thermodynamics.
In this section  we also analyze the thermodynamic properties in the Einstein limit of the HL gravity, 
and the special case of a black hole with flat horizon.  
Finally, in Sec. \ref{sec:con} we present the conclusions of our work.

\section{Review of geometrothermodynamics}
\label{sec:gtd}

Geometrothermodynamics (GTD) is a theory that has been
formulated recently \cite{quev07} in order to
introduce in a consistent manner the Legendre invariance in 
the geometric description of the space of thermodynamic equilibrium states. 
This theory has been applied to different thermodynamic systems  like black holes, 
the ideal gas or the van der Waals gas \cite{nosotros,hernando4,alejandro,
alejandro2,hernando5,hernando6}. In all the cases analyzed so far, 
GTD has delivered consistent results and allows us to  describe geometrically the thermodynamic 
interaction and the phase transitions by meas of Legendre invariant metrics.

The main ingredient of GTD is a $(2n+1)$--dimensional
manifold $\mathcal{T}$ with coordinates $Z^A=(\Phi,E^a,I^a)$, where 
$\Phi$ is an arbitrary thermodynamic potential, $E^a$, $a=1,2,...,n$, are the
extensive variables, and $I^a$ the intensive variables. It is also possible 
to introduce in a canonical manner the fundamental one--form $\Theta = d\Phi - \delta_{ab}I^a d E^b$,
$\delta_{ab}={\rm diag}(+1,...,+1)$,  
which satisfies the condition $\Theta \wedge (d\Theta)^n \neq 0$, where $n$ is 
the number of thermodynamic degrees of freedom of the system, and is invariant
with respect to Legendre transformations 
$(\Phi,E^{a},I^{a})\rightarrow(\tilde{\Phi},\tilde{E^{a}},\tilde{I^{a}})$ with
$ \Phi = \tilde \Phi - \delta_{ab} \tilde E ^a \tilde {I^b},$ 
$ E^a = - \tilde I ^ {a},$ and $  I^{a} = \tilde E ^ a$. Moreover, we assume that on ${\cal T}$ there
exists a metric $G$ which is also invariant with respect to Legendre transformations. 
The triad  $({\cal T},\Theta,G)$ defines a Riemannian contact manifold which is called 
the thermodynamic phase space (phase manifold). The space of thermodynamic equilibrium states 
(equilibrium manifold)  is an $n-$dimensional
Riemannian submanifold ${\cal E}\subset {\cal T}$ induced by a smooth map
$\varphi:{\cal E}\rightarrow{\cal  T}$ , i.e. $\varphi:(E^a) \mapsto
(\Phi,E^a,I^a)$, with $\Phi=\Phi(E^a)$ and $I^a= I^a(E^a)$,  such that
$ \varphi^*(\Theta)=\varphi^*(d\Phi - \delta_{ab}I^{a}dE^{b})=0 $
holds, where $\varphi^*$ is the pullback of $\varphi$. The manifold ${\cal E}$ is 
naturally equipped with the Riemannian metric $g=\varphi^*(G)$. The purpose of GTD is to 
demonstrate that the geometric properties of ${\cal E}$ are related to the thermodynamic properties
of a system with fundamental thermodynamic equation $\Phi=\Phi(E^a)$.

The nondegenerate Legendre invariant metric \cite{qstv10a}
\begin{equation}
G = \Theta^2  +
(\chi_{ab}E^{a}I^{b})(\eta_{cd}dE^{c}dI^{d}) ,\  \eta_{ab} =
{\rm diag}(-1, 1, ..., 1) \ ,
\label{gup1}
\end{equation}
where $\chi_{ab}$ is a diagonal constant tensor, has been used extensively 
to describe second order phase transitions, especially in the context of black hole thermodynamics. 
The arbitrariness contained in the choice of the constant tensor $\chi_{ab}$ has been used to 
simplify the final form of the metric $g=\varphi^*(G)$. For instance, if $\chi_{ab}=\delta_{ab}$, the term
$\varphi^*(\delta_{ab}E^{a}I^{b})$ turns out to be proportional to the thermodynamic potential $\Phi(E^a)$, 
by virtue of the Euler  identity \cite{qstv10a}. However, it is also possible to choose $\chi_{ab}=\eta_{ab}$ without 
affecting the Legendre invariance. In this way, we found recently \cite{mexpak10,tq11} that a slightly generalized metric 
\begin{equation}
G=\Theta^{2}+\frac{1}{2}\left[(\delta_{ab}-\eta_{ab}) E^{a}I^{b}\right]
\left(\eta_{cd}dE^{c}dI^{d}\right)\ ,
\label{gupgen}
\end{equation}
can be used to handle in a geometric manner not only second order phase transitions, but also the thermodynamic limit $T\rightarrow 0$.   
In this work, we consider the metric (\ref{gupgen}) that induces on ${\cal E}$, by means of $g
= \varphi^{\ast}(G)$, the thermodynamic metric
\begin{equation}
g=\frac{1}{2}\left[  E^{a}\left( \frac{\partial{\Phi}}{\partial{E^{a}}} - 
\eta_{ab}\delta^{bc}\frac{\partial{\Phi}}{\partial{E^{c}}}\right)\right]
\left(\eta
_{ab}\delta^{bc}\frac{\partial^{2}\Phi}{\partial {E^{c}}\partial{E^{d}}}
dE^a dE ^d
\right).
\label{gdown}
\end{equation}

Note that in the formalism of GTD the metric
\be 
g_0= \frac{\partial^2\Phi}{\partial E^a\partial E^b} dE^a dE ^b 
\ee
is generated as $g_0=\varphi(G_0)=\varphi^*(\delta_{ab} d E^a dI^b)$, where the metric $G_0$ is not Legendre invariant. This implies that 
the results obtained by using the metric $g_0$ can depend on the choice of thermodynamic potential and, consequently, can lead to 
contradictory results. In particular, for $\Phi = U$ (internal energy), we obtain that $g_0$ is equivalent to the Weinhold metric, 
and for $\Phi=S$, $g_0$ is the Ruppeiner metric. If $\Phi$ is any thermodynamic potential that can be obtained from $U$ 
by means of a Legendre transformation, $g_0$ turns out to be proportional to the metric $g^{LLLS}$ given in Eq.(\ref{llls}).


\section{Topological black hole solutions in HL gravity}
\label{sec:hor}

The HL gravity breaks general 4D covariance and splits it into 3D covariance plus 
reparametrization invariance of time. It is therefore convenient to formulate it 
in the $(3+1)$--ADM formalism, where an arbitrary  metric can be
written in the form
\be 
\label{metricaADM} 
ds^2=-N^2dt^2+g_{ij}(dx^i+N^i dt)(dx^j
+N^jdt)\,,
\ee 
where $N^2$ is the lapse function and $N^i$ represents the shift. Then, the HL action 
is written as   \cite{Lu},
\be
 \label{horava1} 
 \mathcal{I}_{HL}=\int{\mathcal{L}_{HL}\,\,
dt\,\, d^3 x }\,,\ee 
where
\bea 
\label{horava2} \mathcal{L}_{HL}=\sqrt{g} N &\Bigg[&
\frac{2}{\kappa} \Big( K_{ij}K^{ij}-\lambda K^2 \Big)+
\frac{\kappa^2\mu^2(\Lambda R -3\Lambda^2)}{8(1-3\lambda)}+
\frac{\kappa^2\mu^2(1-4\Lambda) }{32(1-3\lambda)} R^2- \nonumber
\\ &-&
 \frac{\kappa^2}{2\omega^4} \Bigg( C_{ij}-\frac{\mu
\omega^2}{2} R_{ij} \Bigg)\Bigg( C^{ij}-\frac{\mu \omega^2}{2}
R^{ij} \Bigg) \Bigg]\,.\eea 
Here $R_{ij}$ and $R$ are the 3D Ricci tensor and curvature scalar, respectively.
Moreover, the extrinsic curvature $K_{ij}$ and the Cotton tensor $C_{ij}$
are given by the expressions
\be
K_{ij}=\frac{1}{2N}(\dot{g}_{ij}-\nabla_i N_j-\nabla_j N_i)\ ,\quad
C^{ij}=\epsilon^{ikl}\nabla_k
\Bigg(R^j{}_l-\frac{1}{4}\,\,R\,\,\delta^j{}_l \Bigg)\,,
\label{horava3} 
\ee
where a dot represents differentiation with respect to the time coordinate.
Finally, $\kappa^2$, $\lambda$, $\mu$, $\omega$ and $\Lambda$ are
constants parameters. 

As mentioned in Section \ref{sec:int}, the vacuum of this theory turns out to be the anti--de Sitter spacetime;
however, it is possible to consider an additional term ($\mu^4 R)$ in the original action to obtain a Minkowski 
vacuum in the IR limit. This generalization is known as the deformed HL model.

A comparison of the HL action with the Einstein-Hilbert action leads to the conclusion that the speed of light, 
Newton's constant and the cosmological constant $\tilde \Lambda$ are given by 
\be
c=\frac{\kappa^2\mu}{4} \sqrt{\frac{\Lambda}{1-3\lambda}}\ ,\quad G= \frac{\kappa^2 c}{32\pi}\ ,\quad
\tilde\Lambda = \frac{3}{2} \Lambda\ .
\ee

Consider now the spherically symmetric line element 
\be 
\label{horava4} 
ds^2=-\tilde{N}^2(r) f(r)
dt^2+\frac{dr^2}{f(r)}+r^2d\Omega_k^2\,,\ee 
where $d\Omega_k^2$
is the line element of the 2--dimensional Einstein space with
constant curvature $2k$. Substituting the metric (\ref{horava4})
into the action (\ref{horava1}), we obtain \cite{Cai},
\be 
\label{horava5} 
\mathcal{I}_{HL}=\frac{\kappa^2 \mu^2
\Omega_k
}{8(1-3\lambda)}\int{\tilde{N}\Bigg[\frac{(\lambda-1)}{2}F^{\prime
2}-\frac{2\lambda}{r}FF^{\prime}+\frac{(2\lambda-1)}{r^2}F^2
\Bigg]\,\, dt\,\, dr }\,,\ee 
where a prime denotes the derivative
with respect to $r$, and $F$ is defined as,
\be 
\label{horava6} 
F(r)=k-\Lambda r^2-f(r)\,.\ee
The variation
of (\ref{horava5}) leads to the following set of equations
\bea 
\label{horava7}
\Bigg(\frac{2\lambda}{r}F-(\lambda-1)F^{\prime}
\Bigg)\tilde{N}^\prime+(\lambda-1)\Bigg( \frac{2}{r^2}F-F^{\prime
\prime} \Bigg)\tilde{N}&=&0\ , \\ (\lambda -1) r^2 F^{\prime
2}-4\lambda r F F^\prime +2(2\lambda-1)F^2&=&0\,,
\eea 
whose solution is
\be 
\label{horava8} 
F(r)=\alpha r^s\,, \quad \tilde{N}=\gamma r^{1-2s}\,, 
\ee 
where  $\alpha$ and $\gamma$ are integration constants and $s$ is given by
\be 
\label{horava9} 
s=\frac{2\lambda -
\sqrt{2(3\lambda-1)}}{\lambda-1}\,. 
\ee
This solution was obtained recently by Cai, Cao and Ohta (CCO) \cite{Cai2}. 
In general, the value of $s$ with a positive sign in front of the square root is also a solution of the above equations. However, in this 
case the asymptotic properties of the solution are not compatible with the properties of a black hole spacetime. 
In the allowed interval $\lambda > 1/3$, i. e. for $s\in (-1,2)$, 
the above solution is asymptotically anti--de Sitter and describes the gravitational field of 
a static black hole.  

In order to obtain the thermodynamic variables of the CCO black hole, 
it is necessary to use the canonical 
Hamilton formulation for the corresponding 
thermodynamic ensemble \cite{Cai}. According to this Hamiltonian
approach, the mass of the black hole is
\be
\label{HT1} 
 M= \frac{c^3\gamma \Omega_k l^{2-2s}}{16\pi G}
\left( \frac{1+s}{2-s}\right)
\left[ \frac{k+\frac{r_+^2}{l^2}}{(\frac{r_+}{l})^s}\right]^2\,,
\ee 
where $l^2 = - 1/\Lambda$ represents the radius of curvature. Moreover, the
Hawking temperature is given by  
\be 
\label{HT3} T=\frac{\gamma}{4\pi r_+^{2s}}
\left[(2-s)\frac{r_+^2}{l^2} -ks\right]\,.
\ee 
Finally, integrating the first law  of thermodynamics, $dM=TdS+\mu_i d Q^i$, for constant values of the 
additional thermodynamic variables $Q^i$,  the entropy associated with
the black hole is obtained as
\be
\label{HT2} 
S=\frac{c^3 \Omega_k l^{2}}{4 G} \left(  
\frac{1+s}{2-s}\right)\left( \frac{r_+^2}{l^2} + 
k\ln \frac{r_+^2}{l^2} \right)\,.
\ee
The entropy $S$ is defined up to an additive constant that can be chose arbitrarily in order to avoid zero or negative values.
Here, $r_+$ represents the radius of the exterior horizon 
which is a function  of $M$ and $l$ determined by the
algebraic equation
\be
\label{HT4} 
r_+^s- \frac{\mathbb{A}}{M^{\frac{1}{2}}
l^2} r_+^2-\frac{\mathbb{A}k}{M^{\frac{1}{2}}}=0\,, \quad \mathbb{A}=\frac{ \kappa \mu
\gamma^{\frac{1}{2}}\Omega_k^{\frac{1}{2}}
}{2^{\frac{7}{4}}[3\lambda -1]^{\frac{1}{4}}}\ .
\ee
Using the expressions (\ref{HT1}) and (\ref{HT3}), we obtain  the heat capacity
\be
\label{HT61} 
C=\Bigg(\frac{\partial M}{\partial
r_+}\Bigg)\Bigg(\frac{\partial T}{\partial
r_+}\Bigg)^{-1}=
\frac{c^3\Omega_k l^{2}}{4 G} \Bigg(
\frac{1+s}{2-s}\Bigg)\left[
\frac{\left(k+\frac{r_+^2}{l^2}\right)\left[(2-s)\frac{r_+^2}{l^2}-ks
\right]}{(s-1)(s-2)\frac{r_+^2}{l^2}+ks^2 }\right]\,.
\ee

According to Davies \cite{davies}, second order phase transitions take place 
at those points where the heat capacity diverges, i. e., for
\be
\frac{r_+^2}{l^2} = \frac{ks^2}{(s-1)(2-s)}\ .
\label{phases}
\ee
Then, we conclude that phase transitions can occur only for $k=1$ and  $s\in (1,2)$,
and for $k=-1$ and $s\in (-1,1)$. For all the remaining values of $k$ and $s$ the 
corresponding black hole cannot undergo a phase transition. Notice, however, that the
phase transition condition (\ref{phases}) must be considered together with
the inequality 
\be
\frac{r_+^2}{l^2}> \frac{ks}{2-s}
\ee
that follows from the condition $T>0$ from Eq.(\ref{HT3}).


\section{Weinhold and Ruppeiner geometries}
\label{sec:geo}

According to Eqs.(\ref{HT1})--(\ref{HT2}), the mass of the CCO black hole is a function of
the entropy  $S$ and the curvature radius $l$. Although the entropy is clearly a thermodynamic 
variable, the thermodynamic nature of the radius of curvature is not so obvious. Nevertheless,
a detailed analysis \cite{cal00} of the thermodynamic properties of AdS black holes reveals that indeed 
it is possible to consider the cosmological constant as a well-defined thermodynamic variable.
Here, we follow this result and assume that the radius of curvature is a thermodynamic variable.

\begin{figure}
\includegraphics[width=5cm]{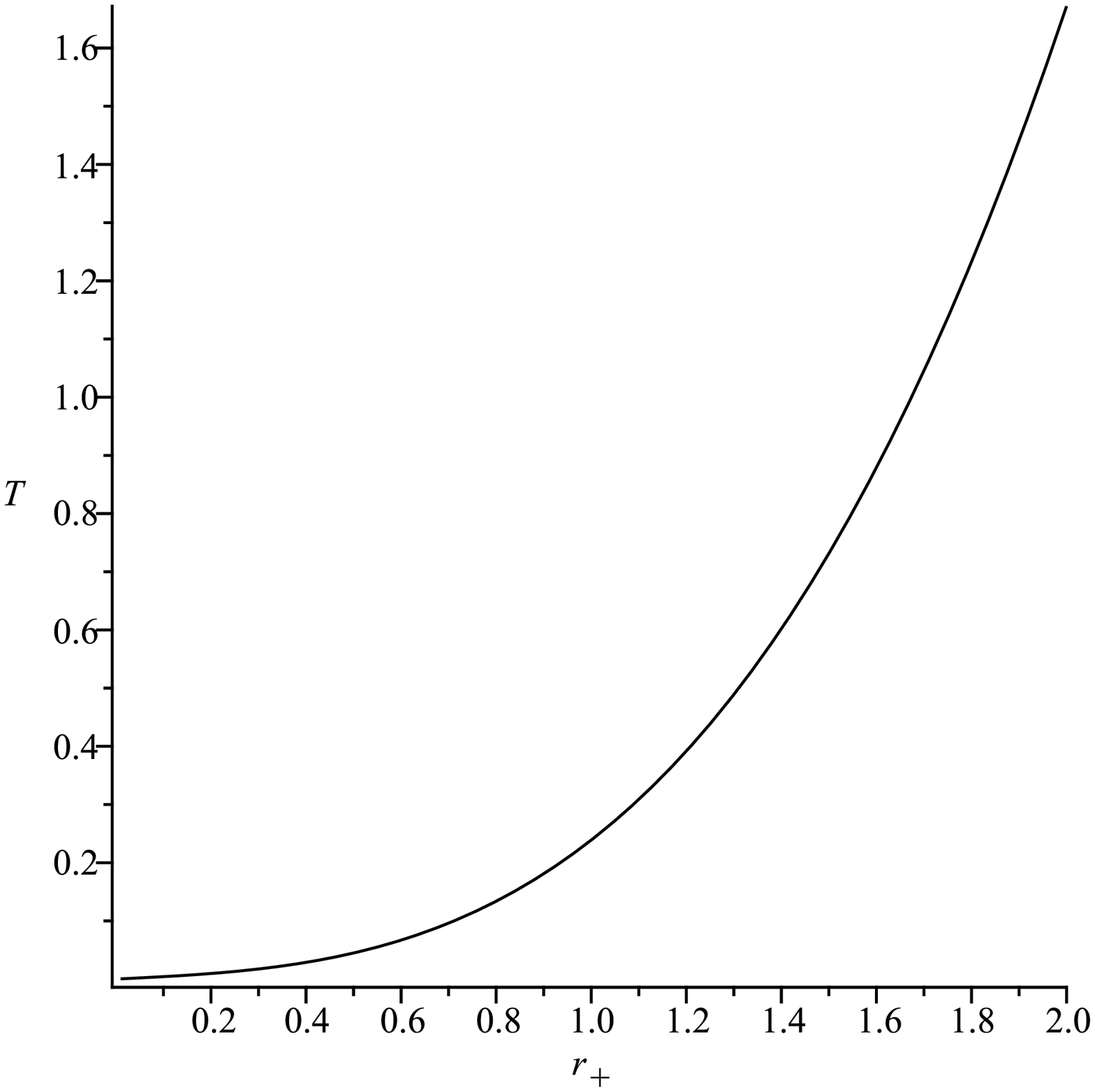}
\includegraphics[width=5cm]{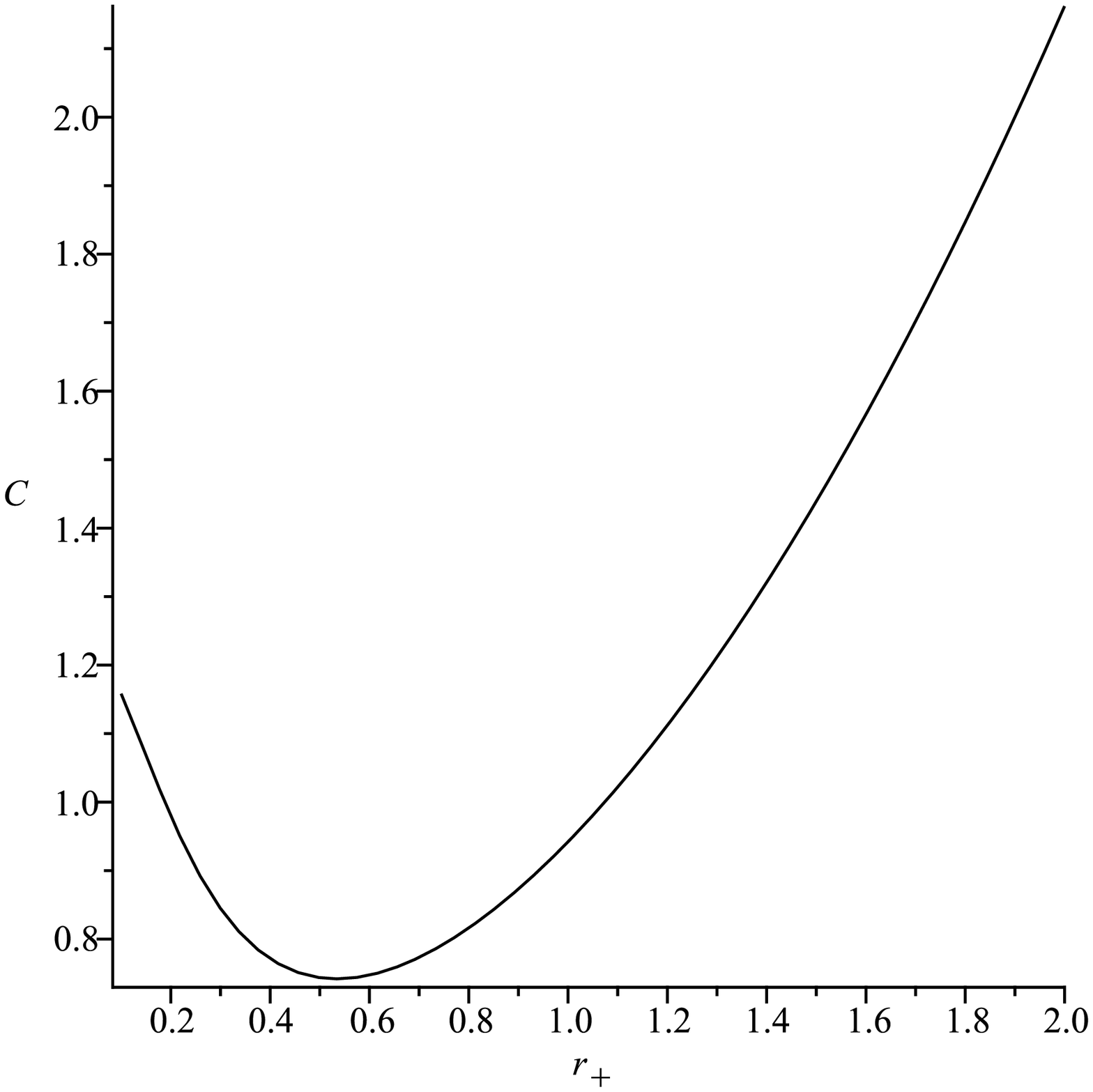}
\includegraphics[width=5cm]{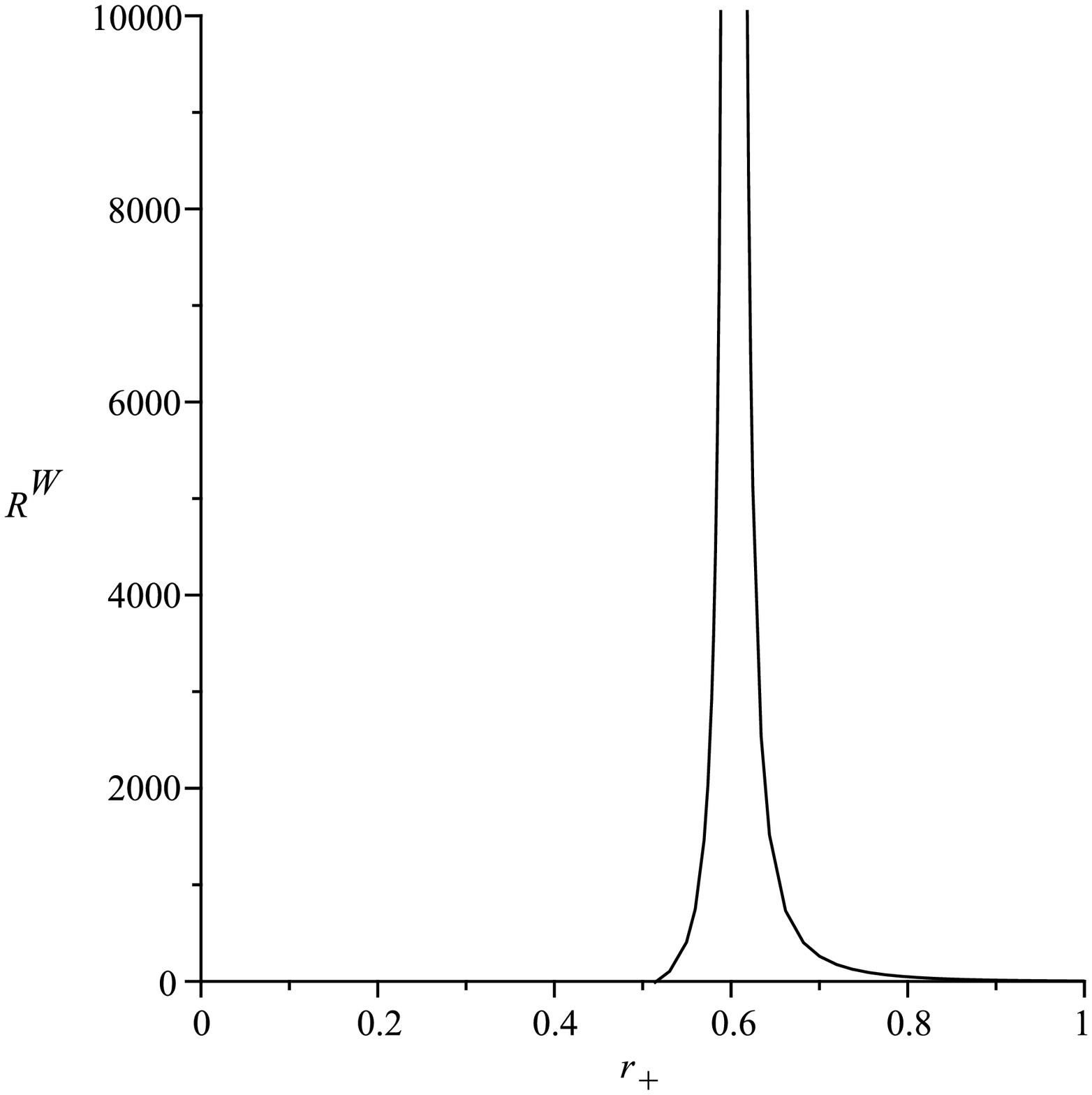}
\caption{Temperature, heat capacity, and Weinhold curvature of the topological black hole. 
We choose $k=1$, $l=1$ and $s=-1/2$ so that $T>0$ for $r_+>0$. 
Moreover, we take $\gamma=1$ and $\Omega_kc^3/(4G)=1$ for simplicity. The heat capacity is a smooth function 
in the entire region $r_+>0$ with a minimum located at $r_+\approx 0.53$. 
The curvature shows a singularity at $r_+\approx 0.60$.}
\label{fig1}
\end{figure}

Let us consider first the Weinhold metric Eq.(\ref{wei}). Since in the case of black holes 
the internal energy is represented by the mass $M$, the Weinhold metric becomes
\be
g^W = M_{SS} dS^2 + 2 M_{Sl} dS dl + M_{ll} dl^2 \ ,
\ee
where $M_S = \partial M /\partial S$, etc. Since the expression for $M$ as given in Eq.(\ref{HT1}), does not 
contain $S$ explicitly, it is necessary to use $r_+$ as a coordinate and its relation to $S$ and $l$ by means of 
Eqs.(\ref{HT3}) and (\ref{HT2}). Then, we obtain
\be
g^W = M_{SS} S_{r_+}^2 dr_+^2 + 2 \left( M_{SS}S_l + M_{Sl} \right)S_{r_+} dr_+ d l + 
\left(M_{SS}S_l^2 + 2 M_{Sl}S_l + M_{ll}\right) dl^2\ .
\ee
Introducing the thermodynamic equations (\ref{HT1})--(\ref{HT2}), we obtain the explicit metric components
\be
g^W_{r_{_+}r_{_+}} = \frac{c^3\Omega_k\gamma}{4\pi G} \left(\frac{1+s}{2-s}\right) l^{4-2s}\left(\frac{r_+}{l}\right)^{-2-2s}\left(k + \frac{r_+^2}{l^2}\right)
\left[ks^2 + (s^2-3s+2)\frac{r_+^2}{l^2} \right]\ ,
\ee
\bea
g_{lr_{_+}}=&&\frac{c^3\Omega_k\gamma}{4\pi G r_+}\left(\frac{1+s}{2-s}\right) l^{-2s} \left(\frac{r_+}{l}\right)^{-1-2s}
\Bigg\{k\left[ks^2+ (s^2-3s+2)\frac{r_+^2}{l^2} \right]\ln\frac{r_+^2}{l^2}  
\nonumber \\
& & + (s-2)\frac{r_+^2}{l^2} \left[\frac{r_+^2}{l^2}-k(s-2)\right] - k^2s^2 \Bigg\}\ ,
\eea
\bea
g^W_{ll}= && \frac{c^3\Omega_k\gamma}{8 \pi G(k+r_+^2/l^2)r_+^{2s}}\left(\frac{1+s}{2-s}\right) 
\Bigg\{ 2k^2 \left[ ks^2+(s^2-3s+2)\frac{r_+^2}{l^2} \right]\ln^2\frac{r_+^2}{l^2}
\nonumber\\
&& + 4 k  \left[(s-2)\frac{r_+^2}{l^2}\left[\frac{r_+^2}{l^2}-k(s-2)\right] - k^2s^2 \right] \ln\frac{r_+^2}{l^2} \nonumber\\
&& + 3\frac{r_+^6}{l^6} + k(11-4s)\frac{r_+^4}{l^4} + k^2(2s^2-10s+13)\frac{r_+^2}{l^2} + k^3(1+2s^2)\Bigg\} \ .
\eea
Moreover, the curvature scalar can be expressed as  
\be
R^W = \frac{N^W}{D^W}\ ,\quad D^W = (s^2-s-2)r_+^6 + k(s^2+8s-8)l^2r_+^4 + k^2(s^2-3s+2)l^4r_+^2+k^3s^2l^6\ ,
\ee
where $N^W$ is a function of $r_+$ and $l$ which is finite at those points where the denominator vanishes.  
The singularities are determined by the roots of the equation $D^W=0$. It is easy to see that the solutions 
of this equation do not coincide with the points where the heat capacity diverges. Figure \ref{fig1} shows 
the concrete example of a stable black hole in which a curvature singularity exists at a point 
where the heat capacity is regular. We conclude 
that the Weinhold curvature fails to reproduce the phase transition structure of the CCO topological black hole.

We now consider the Ruppeiner metric
\be
 g^R= S_{MM} dM^2 + 2S_{Ml} dM dl + S_{ll} dl^2 \ , 
 \ee
which in terms of the coordinates $r_+$ and $l$ becomes
\be
g^R = S_{MM}M_{r_{_+}}^2 dr_+^2 + 2 \left(S_{Ml} + S_{MM} M_l\right) M_{r_{_+}} dr_+ dl 
+ \left(S_{MM} M_l^2 + 2 S_{Ml} M_l + S_{ll} \right)dl^2 \ .
\ee
Using the expressions for the thermodynamic variables (\ref{HT1}) and (\ref{HT2}), we obtain
\be
g^R_{r_{_+}r_{_+}}=- \frac{c^3\Omega_k}{G}\left(\frac{1+s}{2-s}\right)
 \frac{l^2}{r_{_+}^2} \Big(k+\frac{r_{_+}^2}{l^2}
\Big)\frac{\Big[s^2\Big(k+\frac{r_{_+}^2}{l^2}
\Big)+(2-3s)\frac{r_{_+}^2}{l^2} \Big]}{\Big[s\Big(k+\frac{r_{_+}^2}{l^2}
\Big)-2\frac{r_{_+}^2}{l^2} \Big]}\,,
\ee
\be 
g^R_{lr_{_+}}=-\frac{c^3\Omega_k}{G}\left(\frac{1+s}{2-s}\right)
 \frac{l}{r_{_+}} \Big(k+\frac{r_{_+}^2}{l^2}
\Big)\frac{\Big[s^2k\Big(k-\frac{r_{_+}^2}{l^2}
\Big)-k(s-2)\frac{r_{_+}^2}{l^2}-(2s^2-7s+6)\frac{r_{_+}^4}{l^4}
\Big]}{\Big[s\Big(k+\frac{r_{_+}^2}{l^2} \Big)-2\frac{r_{_+}^2}{l^2}
\Big]^2}\,,
\ee
\bea
g^R_{ll}&=&\frac{c^3\Omega_k}{2G}\left(\frac{1+s}{2-s}\right)
\frac{1} {\Big[s\Big(k+\frac{r_{_+}^2}{l^2} \Big)-2\frac{r_{_+}^2}{l^2} \Big]^3}
\Bigg\{
2ks\frac{r_{_+}^2}{l^2}\Big[(11s-19)\frac{r_{_+}^4}{l^4}+7k(2s-1)\frac{r_{_+}^2}{l^2}+k^2\Big] \nonumber \\
&+& k\ln{\Big(
\frac{r_{_+}^2}{l^2}\Big)}\Big[(s^3-6s^2+12s-8)\frac{r_{_+}^6}{l^6}+ks(3s^2-12s+12)\frac{r_{_+}^4}{l^4}+3k^2s^2(s-2)\frac{r_{_+}^2}{l^2}+k^3s^3
\Big]+\nonumber \\ 
&+& k\Bigg[(20-3s^3)\frac{r_{_+}^6}{l^6}-k(20+9s^2)\frac{r_{_+}^4}{l^4}+k^2(4-9s^3)\frac{r_{_+}^2}{l^2}+k^3s^2(2-3s)
\Bigg] +(6s^2-22s+20)\frac{r_{_+}^8}{l^8}\Bigg\}\nonumber\,.
\eea

\begin{figure}
\includegraphics[width=7cm]{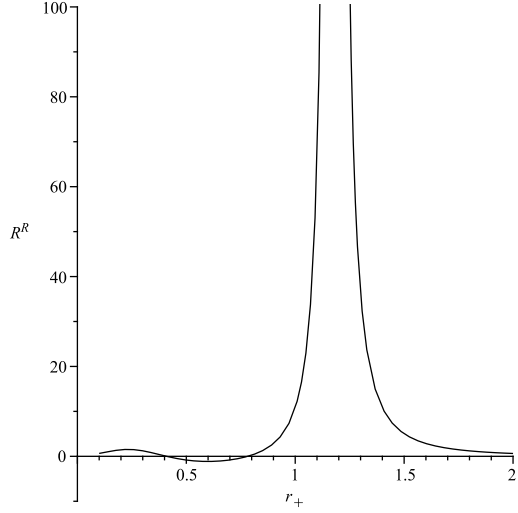}
\caption{Thermodynamic curvature of the Ruppeiner metric. This case corresponds to the choice: $k=1$, $l=1$, $s=-1/2$,
$\gamma=1$, and $\Omega_kc^3/(4G)=1$. A singularity exists at $r_+\approx 1.18$ which does not correspond to a phase 
transition.}
\label{fig2}
\end{figure}

From these expressions for the metric functions it is then straightforward to find the scalar curvature
\be
R^R = \frac{N^R}{D^R}\ ,  \ \ 
D^R =(s+1)l^{10}\Big[A \ln{\Big(\frac{r_{_+}^2}{l^2} \Big)}
-B\Big]^2 \Big(k+\frac{r_{_+}^2}{l^2} \Big)^2
\Big[s\Big(k+\frac{r_{_+}^2}{l^2} \Big)-2\frac{r_{_+}^2}{l^2} \Big]^3\,,
\ee 
with
\be
A=kl^6\Big[\frac{r_{_+}^4}{l^4}(s^3-5s^2+8s-4)+\frac{r_{_+}^2}{l^2}(2s^2-5s+2)ks+k^2s^3
\Big]\,, 
\ee
\be
B=kl^6\Big[\frac{r_{_+}^4}{l^4}(3s^3-13s^2+16s-4)+\frac{r_{_+}^2}{l^2}(6s^2-15s+6)ks+3k^2s^3
\Big]+r_{_+}^6(2s^2-8s+8)\,. \ee
The curvature singularities are determined by the roots of the equation $D^R=0$ which do not coincide with the 
points where the heat capacity (\ref{HT61}) shows second order phase transitions. To illustrate the behavior of 
the curvature we analyze the particular case with $k=1$, $l=1$, and $s=-1/2$ for which the temperature is 
always positive and the heat capacity is a smooth positive function which corresponds to a stable black hole 
configuration. The numerical analysis of this case is depicted in Fig.\ref{fig2}.

\section{Geometrothermodynamics of the CCO black holes}
\label{sec:gtd1}

\begin{figure}
\includegraphics[width=7cm]{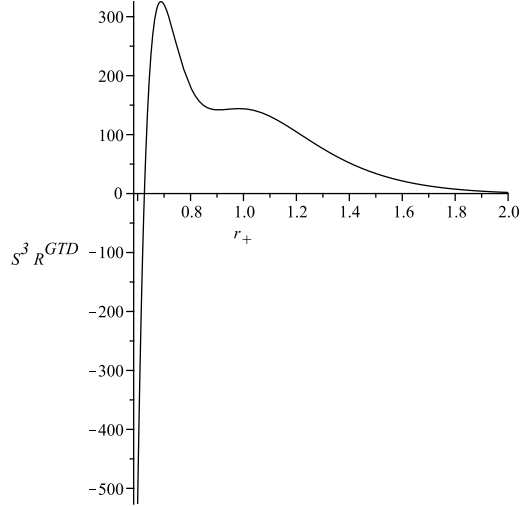}
\caption{Thermodynamic curvature of the GTD metric. This case corresponds to the choice: $k=1$, $l=1$, $s=-1/2$,
$\gamma=1$, and $\Omega_kc^3/(4G)=1$. The expression $S^3 R^{^{GTD}}$ is plotted to avoid the unphysical singularity at $S=0$.
}
\label{fig3}
\end{figure}

The formalism of GTD is invariant and, consequently,  we can choose any arbitrary 
thermodynamic potential $\Phi$ in any arbitrary representation to describe the thermodynamics of a black hole. 
Take, for instance, $\Phi=M$ for the CCO topological black holes presented in Sec. \ref{sec:hor}. The coordinates
of the 5-dimensional phase manifold can be chosen as $Z^A=(M,S,l,T,L)$, where $T$ is the temperature dual to $S$ and
$L$ is the dual of the curvature radius $l$. The fundamental one-form is then $\Theta = dM - TdS -l dL$ and the Legendre
invariant metric (\ref{gupgen}) is written as
\be
G= \Theta^2 + ST \left( - dS dT + dl dL\right)\ . 
\ee
The smooth map $\varphi: {\cal E} \rightarrow {\cal T}$ or in coordinates $\varphi: (S,l) \mapsto [M(S,l), S, l, T(S,l), L(S,l)]$
determines the equilibrium manifold ${\cal E}$ with metric
\be
g^{GTD} = \varphi^*(G)= S M_S \left(-M_{SS} dS^2 + M_{ll} dl^2\right) \ ,
\ee
on which the first law of thermodynamics $dM= TdS + l dL$ and the equilibrium conditions
\be
T=\frac{\partial M}{\partial S} \equiv M_S\ ,\quad L = \frac{\partial M}{\partial l} \equiv M_l 
\ee
hold. As  mentioned above, the fundamental equation $M=M(S,l)$ cannot be written explicitly and so we use 
 $r_+$ instead of $S$ as a coordinate. Then, 
\be
g^{GTD} = SM_S\left[ - M_{SS}S_{r_{+}}^2 dr_+^2 - 2M_{SS} S_{r_{+}} S_l dl dr_+ + (M_{ll} - M_{SS}S_l^2)dl^2\right]\ .
\label{gdownr} 
\ee
Using the expressions for the mass and the entropy, we obtain
\bea
g^{GTD}&=& -\frac{c^6\Omega_k^2\gamma^2l^6r_+^{4s}}{8\pi G}\left(\frac{s+1}{2-s}\right) 
\left[(s-2)\frac{r_+^2}{l^2} + ks\right] S(r_+,l)  \nonumber\\
&& \times\Bigg\{\left[(s-1)(s-2)\frac{r_+^2}{l^2} + k s^2\right] 
\left[ dr_+^2 + \frac{2r_+}{l\left( k + {r_+^2}/{l^2}\right)}dldr_+\right] + C(r_+,l) dl^2\Bigg\} \ ,
\eea
where
\be
C(r_+,l) = \frac{ 16k^2 \frac{r_+^3}{l^3} \left( \ln \frac{r_+^2}{l^2}\, -1\right)^2\left[(s-1)(s-2)\frac{r_+^2}{l^2} + ks^2\right]
- \left( k^2 + 3 \frac{r_+^4}{l^4} \right)\left( k + \frac{r_+^2}{l^2}\right) }{16l^2 \left( k + \frac{r_+^2}{l^2}\right) }
\ee
The curvature scalar corresponding to the metric (\ref{gdownr}) is found to be   
\bea
R^{^{GTD}}=\frac{N^{^{GTD}}}{D^{^{GTD}}}\ ,\quad  
D^{^{GTD}}&=&  \left( {k}^{2}+3\,\frac{r_{_+}^{4}}{l^4}  \right) ^{2}
\left( k+{\frac {r_{_+}^{2}}{{l}^{2}}} \right) ^{13} \left( {\frac {r_{_+}^
{2}}{{l}^{2}}}+k\ln   {\frac {r_{_+}^{2}}{{l}^{2}}}  
 \right) ^{3} \nonumber \\ 
&&\times \left[ { { \left( 2-s \right)} \frac{r_{_+}^{2}}{{{l}^{2}}}}-ks
 \right] ^{4} \left[  { \left( s-1 \right)  \left( s-2 \right)}\frac{ r_{_+}^{2}} {{l}
^{2}}+k{s}^{2} \right] ^{2}
\eea
where $N^{^{GTD}}$ is a function of $r_+$ and $l$ that is finite at those points where the denominator vanishes. 
There are several curvature singularities in this case. The first one occurs if $k+r_+^2/l^2=0$ and corresponds to the 
limit $M\rightarrow 0$, as follows from Eq.(\ref{HT1}). A second singularity is located at 
the roots of the equation $r_+^2+k\ln r_+^2 = 0$ and can be interpreted from Eq.(\ref{HT2}) as the limit $S\rightarrow 0$.
Moreover, according to Eq.(\ref{HT3}), the singularity situated at $r_+^2/l^2= ks/(2-s)$ corresponds  the limit $T\rightarrow 0$.
Finally, if $(s-1)(s-2)r_+^2/l^2+ks^2=0$ a singularity occurs that, according to Eq.(\ref{HT61}),  coincides with the 
limit $C\rightarrow \infty$, i. e., with the points where second order phase transitions take place. Clearly, 
the singularities at which the mass, the entropy or the temperature vanish must be considered as unphysical and indicate
the limit of applicability of the thermodynamics of black holes. The thermodynamic curvature in GTD   for the case $k=1$, $l=1$
and $s=-1/2$ shows a singularity at the value $r_+^2+\ln r_+^2 = 0$, i. e., for $r_+\approx 0.75$, which corresponds to the limit $S\rightarrow 0$.
Figure \ref{fig3} illustrates the behavior of the curvature $S^3 R^{^{GTD}}$ to avoid the unphysical singularity as $S\rightarrow 0$.
We see that in the analyzed interval no curvature singularities appear. This is in accordance with the behavior of the heat capacity 
which in the same interval is free of phase transitions (see Fig.\ref{fig1}). A curvature singularity can be observed for $r_+\rightarrow 0$ 
which indicates the break down of the black hole configuration and, consequently, of its thermodynamics.  

We conclude that the curvature obtained within the formalism
of GTD correctly describes the thermodynamic behavior of topological black hoes in HL gravity.

\subsection{The Einstein limit of the CCO black holes} 
\label{sec:eins}

Einstein's theory of gravity with cosmological constant is obtained from the HL gravity in the limit
$\lambda\rightarrow 1$. The CCO topological black holes reduce in this case to a single black hole configuration with $s=1/2$,
whereas the corresponding  thermodynamic variables are written as
\be
\label{H15} 
M=\frac{c^3 \Omega_k l}{16\pi G}\frac{l}{r_+}
\left({k+\frac{r_+^2}{l^2}}\right)^2 \ ,
\ee
\be 
\label{H16} 
S=\frac{c^3 \Omega_k l^{2}}{4 G}
\Bigg(\frac{r_+^2}{l^2}+ k\ln \frac{r_+^2}{l^2} \Bigg) \,,
\ee
\be 
\label{H17} 
T=\frac{1}{8\pi r_+} \Bigg(3\frac{r_+^2}{l^2}-k\Bigg)\, ,
\ee
and 
\be
\label{H18} 
C=\Bigg(\frac{\partial M}{\partial
r_+}\Bigg)\Bigg(\frac{\partial T}{\partial
r_+}\Bigg)^{-1}=
\frac{c^3\Omega_k l^2}{2
G}\frac{\Big(k+\frac{r_+^2}{l^2} \Big)\Big(3\frac{r_+^2}{l^2} - k \Big)}
{\Big(k+3\frac{r_+^2}{l^2} \Big)} \,.
\ee
It follows that the second order phase transitions take place at the points where the condition $3r_+^2/l^2 + k =0$ is satisfied.

Introducing the expressions (\ref{H15}) and (\ref{H16}) into the metric (\ref{gdownr}), we obtain
\be
 \label{H19} 
 g^{GTD} = \frac{1}{32}\left(\frac{c^3\Omega_k \gamma}{4 G}\right)^2 \left(\frac{r_+^2}{l^2} + k \ln \frac{r_+^2}{l^2}\right) \left(k - 3\frac{r_+^2}{l^2}\right)
\frac{l^2}{r_+^2} \left( A_1 dr_+^2 + 2 A_2 dl dr_+  + A_3 dl^2\right)\ ,
 \ee
with
\be
A_1 = \frac{l^2}{r_+^2} \left(k + \frac{r_+^2}{l^2}\right)\left(k+3\frac{r_+^2}{l^2}\right) \ ,
\ee
\be
A_2 = k \frac{l}{r_+} \left(\ln\frac{r_+^2}{l^2}\, - 1\right) \left(k+3\frac{r_+^2}{l^2}\right) \ , 
\ee
\be
A_3 = \frac{1}{k+\frac{r_+^2}{l^2}}\left[k^2\left(k + 3 \frac{r_+^2}{l^2}\right)  \left(\ln \frac{r_+^2}{l^2}\, - 2\right)\ln \frac{r_+^2}{l^2} 
+ 6\frac{r_+^4}{l^4}\left(k+\frac{r_+^2}{l^2}\right) + k^2\left(k-\frac{r_+^2}{l^2}\right)\right]\ .
\ee
The corresponding scalar curvature can be expressed as
\be
R^{GTD} =\frac{N^{GTD}}{D^{GTD}} \ ,
\ee
\be
D^{GTD} = \left( k+{\frac {{r}^{2}}{{l}^{2}}} \right) ^{4} \left( {\frac {{r}^{
2}}{{l}^{2}}}+k\ln   {\frac {{r}^{2}}{{l}^{2}}}  
 \right) ^{3} \left( 3\,{\frac {{r}^{2}}{{l}^{2}}}-k \right) ^{3}
 \left( 3\,{\frac {{r}^{4}}{{l}^{4}}}+{k}^{2} \right) ^{2} \left( 3\,{
\frac {{r}^{2}}{{l}^{2}}}+k \right) ^{2}
\ ,
\ee
where the numerator function $N^{GTD}$ is finite at all the points where the denominator vanishes. From this expression
we can see that the roots of the equation  $3r_+^2/l^2 + k =0$ determine curvature singularities which coincide with 
the points where second order phase transitions occur $(C\rightarrow \infty)$. Additional singularities occur if
$r_+^2/l^2 + k =0$, $r_+^2/l^2 + k \ln (r_+^2/l^2)=0$, or $3r_+^2/l^2 - k =0$ which correspond to the limits 
$M\rightarrow 0$, $S\rightarrow 0$ or $T\rightarrow 0$, respectively.

\subsection{The limiting black hole with flat horizon}
\label{sec:flat}

According to \cite{Cai},  the thermodynamics of a CCO black hole with a flat horizon ($k=0$) 
is described the following variables 
\be
\label{HF1} 
M=\frac{c^3\gamma \Omega_k l^{2-2s}}{16\pi G}
\Bigg(
\frac{1+s}{2-s}\Bigg)\left(\frac{r_+}{l}\right) ^{2(2-s)}\,,
\ee
\be 
\label{HF2} 
S=\frac{c^3 \Omega_k l^{2}}{4 G} \Bigg(
\frac{1+s}{2-s}\Bigg)\left(\frac{r_+}{l}\right)^2 \,,
\ee
\be 
\label{HF3} 
T=\frac{\gamma l^{-2s}}{4\pi}(2-s)
\left(\frac{r_+}{l}\right)^{2-2s}\,,
\ee
and 
\be 
\label{HF4} 
C=\Bigg(\frac{\partial M}{\partial
r_+}\Bigg)\Bigg(\frac{\partial T}{\partial
r_+}\Bigg)^{-1}=\frac{c^3\Omega_k }{4 G} \Bigg(
\frac{1+s}{2-s}\Bigg)\frac{r_+^2}{s-1}=\frac{S}{s-1}\,.
\ee
From the expression for the heat capacity we see that this black hole is free of phase transitions. 

Using the relations (\ref{HF1}) and (\ref{HF2}), the thermodynamic metric (\ref{gdownr}) is written as
\be
\label{H7} 
g^{GTD}=\frac{s r_+^6}{8}\Bigg(\frac{c^3\gamma
\Omega_k}{4\pi G} \Bigg)^2\Bigg(\frac{1+s}{s-2}
\Bigg)^2\Bigg(\frac{r_+}{l} \Bigg)^{-4s}
\left[ 2(s-1)(s-2)l^{-4(s+1)}dr_+^2-3l^{-2(3+2s)}r_+^2dl^2\right] \,,
\ee
for which we find that the curvature $R=0$, indicating no phase transition structure exists. This is in accordance
with the result obtained above from the study of the heat capacity.   

According to GTD, a flat equilibrium manifold is a consequence of the lack 
of thermodynamic interaction. This can be understood in the following way. For this special case, Eq.(\ref{HF2}) indicates 
that the horizon radius is 
\be
\label{H11} 
r_+=\Bigg[ \frac{4G}{\Omega_k c^3}\Bigg(\frac{2-s}{1+s} \Bigg)S\Bigg]^{\frac{1}{2}}\,,
\ee
so that Eq.(\ref{HF1}) generates  the explicit fundamental equation
\be 
\label{H12} 
M=\frac{\gamma}{4\pi}\Bigg(\frac{4G}{\Omega_k
c^3} \Bigg)^{1-s}\Bigg(\frac{2-s}{1+s}
\Bigg)^{1-s}\frac{S^{2-s}}{l^2}\,.
\ee
which, in turn, can be rewritten as
\be
(2-s)\ln S = \ln M + \ln l^2 + \ln S_0 \ , \quad S_0 = \frac{4\pi}{\gamma}\left(\frac{c^3\Omega_0}{4G}\frac{1+s}{2-s}\right)^{1-s} \ .
\ee
So we see that the entropy function can be separated in the extensive variables $M$ and $l$. 
On the other hand, all thermodynamic potentials 
that possess the property of being separable have been shown \cite{alejandro} to correspond to systems with no thermodynamic interaction and zero 
thermodynamic curvature. This is an indication that the statistical internal structure of a CCO black hole with flat horizon is equivalent
to that of an ideal gas which is the main example of a system with no intrinsic thermodynamic interaction.
We note that in this limiting case Weinhold and Ruppeiner geometries are flat too, indicating also that there exists 
a statistical analogy between a black hole with flat horizon and an ideal gas.



\section{Conclusions}
\label{sec:con}

In this work, we applied the formalism of geometrothermodynamics (GTD) to describe the  
thermodynamics of the Cai-Cao-Ohta (CCO) topological black holes in the Ho\v rava-Lifshitz
model of quantum gravity. In the thermodynamic phase manifold we introduce a particular 
Riemannian metric whose main property is its invariance with respect to Legendre transformations, i. e.,
its geometric characteristics are independent of the choice of thermodynamic potential. This is a 
property which holds in ordinary thermodynamics and we assume as valid in GTD too. The Legendre invariant
metric induces in a canonical manner a thermodynamic metric in the equilibrium manifold which is 
defined as a submanifold of the thermodynamic phase manifold.  

We used the expressions of the main thermodynamic variables of the CCO black holes in order 
to compute the explicit form of the thermodynamic metric of the equilibrium manifold. The 
corresponding thermodynamic curvature turned out to be nonzero in general, indicating 
the presence of thermodynamic interaction. Moreover, it was shown that the 
phase transitions which are characterized by divergencies of the heat capacity 
are described in GTD by curvature singularities in the equilibrium manifold. 
We also studied the thermodynamics of the CCO black holes by using Weinhold and Ruppeiner geometries
and found that they fail to describe the corresponding phase transition structure. These results
are in agreement with a recent work by Janke, Johnston and Kenna \cite{jjk10} in which the 
GTD of the Kehagias-Sfetsos \cite{ks} black hole is investigated. 

It was found that the geometrothermodynamic equilibrium manifold of the COO black holes 
present additional curvature singularities which correspond to the vanishing of the 
mass, entropy and Hawking temperature. We interpret in general the vanishing
of these thermodynamic variables as an indication of the limit of applicability of 
black hole thermodynamics. So we conclude that the formalism of GTD breaks down, with 
curvature singularities, exactly at those points where black hole thermodynamics 
fails. 

In the context of the GTD of the CCO black holes, we also analyzed the limit of Einstein gravity
and of a black holes with flat horizons. In both cases we obtained results which are consistent
with the thermodynamics of the respective black hole configurations. It turned out that 
the equilibrium manifold of
black holes with flat horizons is flat. The flatness of the equilibrium manifold is 
interpreted as a consequence of the  lack of intrinsic thermodynamic interaction. 
This property resembles the statistical behavior of an ideal gas.

\section*{Acknowledgments}

Two of us (HQ and ST) would like to thank ICRANet for support and hospitality,  
and Prof. Asghar Qadir for interesting comments, encouragement and support. 
One of us (ST) gratefully acknowledges a research grant from the Higher
Education Commission of Pakistan under Project No: 20-638. 
This work was supported in part by DGAPA-UNAM, grant No. IN106110.

\end{document}